\documentclass{article}
\usepackage{epsfig}

\usepackage{epsfig}
\usepackage{amsmath}
\usepackage{amsfonts}
\usepackage{amssymb}
\usepackage{graphicx}

\begin{document}

\title{Effects of Static and Dynamic Disorder on the Performance of Neural
Automata}
\author{{J.J. Torres, J. Marro, P.L. Garrido, J.M. Cort\'es, 
F. Ramos, and M.A. Mu\~{n}oz}  \\
{\small Institute \textquotedblleft Carlos I\textquotedblright\ for
Theoretical and Computational Physics, and}\\
{\small \ Departamento de Electromagnetismo y F\'{\i}sica de la Materia,}\\
{\small \ University of Granada, 18071--Granada, Spain}}
\maketitle

\begin{abstract}
We report on both analytical and numerical results concerning stochastic
Hopfield--like neural automata exhibiting the following (biologically
inspired) features: {\bf(1)} Neurons and synapses evolve in time as in contact
with respective baths at different temperatures. 
{\bf(2)} The connectivity
between neurons may be tuned from full connection to high random dilution
or to the case of networks with the small--world property and/or
scale-free  architecture.
{\bf(3)} There is
synaptic kinetics simulating repeated scanning of the stored patterns.
Though these features may apparently result in additional
disorder, the model exhibits, for a wide range of parameter values,
an extraordinary computational performance, and some of the qualitative
behaviors observed in natural systems. In particular, we illustrate here very
efficient and robust associative memory, and jumping between pattern
attractors.
\end{abstract}

\section*{The model and its motivation}

The field of \textit{neural automata,} or cellular automata \cite{pomeau}
that are biologically inspired and aimed at solving open issues in
neuroscience, may be said to initiate with the Hopfield model \cite{hopfield}.
 This illustrates ---on very simple and, consequently, rather unrealistic
grounds--- the property of associative memory, a task for which natural
systems remain far from unbeaten by digital computers. Besides many other
related studies \cite{HopRev}, the Hopfield model has been generalized along
three interesting lines. That is, it has been shown that one may enhance the
computational abilities of a Hopfield--like system by allowing for dynamic
---instead of quenched--- synapses \cite{SN,earlyJJT}, by implementing
dynamics in the computer via parallel ---instead of sequential--- updating 
\cite{little,cortes}, and by considering scale--free ---instead of extremely
connected--- network topologies \cite{SFT,topo}. In this paper we report on
some recent efforts along these lines. We describe a stochastic neural
automata scheme that allows for explicitly studying the influence of the
model details on the resulting computational performance. In particular, we
evaluate the effects of disorder perturbing both (statically) the network
architecture and (dynamically) the synaptic strengths. The importance of
these two details on the functioning of natural neural systems has been recently
recognized, e.g. in Refs.\cite{vitro,joaquinNC} respectively. We
conclude that the computational performance may indeed be importantly
enhanced by appropriately tuning these details. This is illustrated
here with some specific cases; the range of validity of our main results is
known to go beyond these examples, however.

Consider a network with a binary neuron variable, $s_{i}=\pm 1,$ at each
node $i=1,\ldots ,N$. Assume pair synapses of weight $w_{ij}$ connecting the
neurons at $i$ and $j,$ and define the degree $k_{i}$ of node $i$ as the
number of links from $i$ to any other node. A scale--free network \cite{bara}
has power--law distributed degrees, $P\left( k\right) \sim k^{-\gamma },$
while $P\left( k\right) \sim \delta \left( k-N\right) ,$ for the fully
connected case (each neuron links to any other one in the network), and $k$
has mean value $\zeta N,$ $0<\zeta <1,$ for the randomly $\zeta $--diluted
case. The synaptic links are known to serve to store information. In the
standard Hopfield model, for instance, one assumes $M$ patterns consisting
of binary variables, $\xi _{i}^{\mu }=\pm 1,$ $\mu =1,...,M,$ and the
weights are then set according to some \textit{learning rule}. A familiar
instance is the Hebb's rule for which the synapses take values 
$w_{ij}\propto \sum_{\mu}\xi _{i}^{\mu }\xi _{j}^{\mu }.$

The system configuration at time $t$ involves the sets $\mathbf{S}\equiv
\left\{ s_{i}\right\} $ and $\mathbf{W}\equiv \left\{ w_{ij}\right\} $ which
together determine the value of some \textit{energy function.} To be
specific, consider this to be $H\left( \mathbf{S},\mathbf{W};t\right) =-%
\frac{1}{2}\sum_{i}\sum_{j}^{(i)}w_{ij}\left( t\right) s_{i}s_{j}$ where the
second sum goes over all nodes connecting to $i.$ That is, we are assuming
that there is an energy function with the Ising structure \textit{at each
time }$t,$ and that synapses vary with $t.$ For simplicity, we further
assume that $w_{ij}\left( t\right) =\tilde{w}_{ij}\left[ \mu \left( t\right) %
\right] =\sigma N^{-1}\xi _{i}^{\mu \left( t\right) }\xi _{j}^{\mu \left(
t\right) },$ where $\sigma $ is a normalization constant. This amounts to
set each connection at time $t$ equal to the value obtained from the $\mu $
th pattern via a \textit{learning rule.} And our choice
for this (see what follows) is such that, after time averaging, 
it results in the Hebb's rule \cite{note1} (i.e. we average over all patterns
with the same height).
A main fact is that, consistently with the empirical observation that 
memory is a global dynamic phenomenon, we are assuming that all 
local synapses are set
at each time by a given pattern, which is chosen according to some dynamics
to be determined next.

One is mainly interested in $\mathbf{m}\equiv \left\{ m^{\mu };\mu =1,\ldots
,M\right\} $, where $m^{\mu }=m^{\mu }(\mathbf{S})$ is the overlap between
the current state $\mathbf{S}$ and pattern $\mu .$ We assume \cite%
{coolenD,cortes2} that $(\mathbf{S},\mu )$ evolves in discrete time
according to a transition rate  $T$ that may be decomposed  as: 
$T_{\mu ^{\prime }}%
\left[ \mathbf{S}^{\prime }\rightarrow \mathbf{S}\right] \times T_{\mathbf{S}%
}\left[ \mu ^{\prime }\rightarrow \mu \right] ,$ Here, $T_{\mu }$ is taken
as a product of $N$ terms $\Psi \left[ \beta _{0}\Delta H\left(
s_{i}^{\prime }\allowbreak \rightarrow s_{i}\allowbreak =\pm s_{i}^{\prime
}\right) \right] $ (as corresponds to the neuron configuration varying by
parallel updating or \textit{Little dynamics}), and $T_{\mathbf{S}%
}=\allowbreak \Psi \left[ \beta _{1}\Delta H(\mu ^{\prime }\allowbreak
\rightarrow \mu )\right] .$ The function $\Psi \left( X\right) $ is
arbitrary, e.g., of the Glauber type \cite{marro}, and each transition rate
involves both an inverse temperature $\beta $ of the bath inducing its own
stochasticity and the corresponding change $\Delta H$ of the chosen energy
function.

The fact that the neuron activities and the synaptic intensities are
stochastically driven by different parameters, $\beta _{0}^{-1}$ and $\beta
_{1}^{-1},$ respectively ---which one may imagine as the temperatures of two
different baths in contact with each set of degrees of freedom--- endows the
neural automata with a varied and interesting behavior. This is related to
the fact that the competition between baths impedes the system from reaching
thermal equilibrium; in general, the asymptotic state is instead a
nonequilibrium steady state, which is more realistic than an equilibrium
one \cite{marro}. 
Though the baths competition is, to the best of our
knowledge, a novel feature within the present context, our model has some
close antecedents \cite{coolenPRB,joaquinPRA,joaquinPRL}; these, however, 
mainly concern sequential updating, which in general results in a less
efficient mechanism \cite{cortes2}. Furthermore, our model describes the
changes in the two subsystems on the same time scale, which is an
interesting general situation with well--defined limits. That is, $\beta
_{1}\rightarrow \infty $ corresponds to the freezing of synapses \cite%
{hopfield}, large $\beta _{1}$ is for slow synaptic kinetics \cite{coolenPRB}%
, and small $\beta _{1}$ describes extreme synaptic activity \cite%
{earlyJJT,joaquinPRA}.

This neural automata is also simple and versatile enough to allow for
analyzing different network architectures. With this aim, we {\bf (i)}
 first generated
a network starting with $\eta_{0}$ nodes and adding $\eta \leq \eta _{0}$
links at consecutive time steps with the preferential attachment rule.
 This, resulting in $\eta \left( N-\eta_{0}\right)$ 
links for $N$ nodes, will be called the scale--free (Barabasi-Albert) 
network (SFN) \cite{bara}. 
 {\bf (ii)} A fully connected network (FCN) with $N$ nodes has $N^{2}$
links, which is not realistic for a neural system. Instead, {\bf (iii)} 
 a meaningful
alternative to the SFN case is the (highly) diluted network (DN) obtained by
randomly suppressing links in the FCN until only $\eta \left( N-\eta
_{0}\right) $ are left. In order to have a true SFN with the 
\textit{small--world} property, i.e., most of the nodes exhibiting small
connectivity and a few \textit{hubs} having its connectivity comparable to
the network size $N,$ one needs to restrict to $\eta <<N.$

\section*{Scale--free topology of synapses}

The small--world property, which implies that the average path length
between any two nodes is very small compared to the network typical length,
is a suitable feature for an efficient neural system. In fact, this property
has been reported to hold in many natural systems \cite{bara,Mendes},
including growing cultured neurons \cite{vitro}, and it was shown that the
SFN can store and retrieve a given number of patterns with a lower
computer--memory cost that the FCN \cite{SFT}. We shall briefly illustrate
here (see also Ref. \cite{topo}) that the scale--free architecture may
indeed enhance both the network associative performance and its robustness
against thermal noise perturbations.

The graphs in fig.1 show the dependence of the stationary overlap $m^{\mu }$
on the neuron temperature, $T\equiv \beta _{0}^{-1},$ for cases SFN and DN
with $\eta =3$ and $M=1.$ Notice that the SFN with competing temperatures
coincides for $M=1$ with the standard, equilibrium Hopfield model (with
quenched synapses). Fig.1 illustrates that the SFN makes indeed a better job
for retrieving of information than the comparable DN at sufficiently high
temperature, which is the relevant case for practical purposes. As $M$ is
increased, the performance tends to deteriorate in both cases. However, one
then observes an intriguing behavior of the SFN at finite temperature. That
is, there is a definite tendency of the \textit{hubs}, namely, nodes with $%
k\geq k_{0}\left( M,\beta _{0},\beta _{1}\right) ,$ to concentrate most of
the system activity concerning the retrieval process. 

In order to illustrate the above property, we computed the mean 
\textit{local overlap} associated to a node of degree $k,$ $m\left( k\right) $. 
The resulting graphs in fig.2 clearly depict that $m\left( k\right)
\rightarrow 1$ as $k$ increases at any temperature (large fluctuations
simply reflect that the number of hubs is small for the network size used).
This indicates how the hubs, the more the higher its connectivity degree,
tend to become robust references for the process of associative memory. The
state of \textit{boundary nodes}, on the contrary, shows a poor correlation
with the relevant stored pattern. This is in agreement with a previous
observation at zero (neuron) temperature \cite{graw}.
\begin{figure}[h!]
\begin{center}
\epsfxsize=6.5cm
\epsfbox{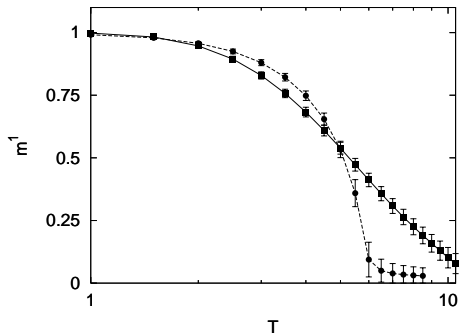}
\caption{Stationary overlap $m^{\mu }$ \textit{versus}
the logarithm of the neuron temperature (arbitrary units), averaged over 100
histories for $N=1600$ neurons, $M=1,$ and $\eta =\eta _{0}=3,$ for the
scale--free network (squares) and for the diluted network (circles). This
illustrates a better performance of the former in general for practical
purposes.}
\end{center}
\end{figure}
The above results triggered our interest in networks in which the connectivity
degree is power--law distributed, $P\left( k\right) \sim k^{-\gamma }.$ An
issue is the possible influence of the parameter $\gamma $ on the system
performance. In fact, the number of highly connected nodes tends to sharply
decrease as $\gamma $ is increased, and it ensues $\gamma \lesssim 2$ as a
convenient range. Further study of this will be reported elsewhere 
\cite{topo}.

\section*{Escaping from the attractor}

Time behavior is also intriguing as illustrated, for instance, by the FCN
architecture, which is more amenable to analysis. Fig.3 depicts the various 
\textit{phases} exhibited by the system as the parameters $\beta _{0}$ and $%
\beta _{1}$ are changed. There is a \textit{ferromagnetic} (F) phase in
which the system shows associative memory, and a \textit{paramagnetic} (P)
phase ---the upper stripe--- lacking this property. This is familiar from
the Hopfield model. A first novelty is that no \textit{mixed states} \cite%
{joaquinPRA} occur, which is computationally convenient. In addition, the
nonequilibrium phase diagram in fig.3 depicts a region between the P and F
phases that exhibits emergent dynamic behavior confirming (and extending) a
result in Ref.\cite{joaquinNC}. That is, the system in this region has 
\textit{dynamic} associative memory, namely, after a transient time in which
one of the stored patterns is recovered, the system jumps to one of the
other possible attractors, and keeps indefinitely doing so.%
\begin{figure}[h!]
\begin{center}
\epsfxsize=6.5cm
\epsfbox{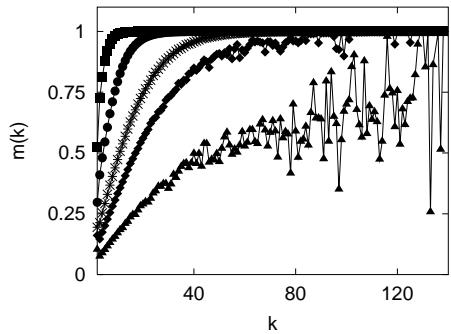}
\caption{The \textit{local overlap,} as defined in the
main text, \textit{versus }the connectivity degree $k$ for different $T$,
increasing from top to bottom, for the scale--free network in fig. 1. This
illustrates a tendency of the \textit{hubs} to control the retrieval of
information.}
\end{center}
\end{figure}
The nature of this jumping deserves a comment. Simulations uncover that
there is some non--trivial structure of time correlations. This is revealed
by monitoring the time $\tau _{\nu ,\zeta }$ the system stays in attractor $%
\nu $ before escaping to another one, say $\zeta .$ Within the largest
jumping region, O(II), $\tau _{\nu ,\zeta }$ happens to be practically
independent of both $\nu $ and $\zeta .$ That is, the system then stays the
same amount of time wandering in each attractor. However, the probability of
jumping between patterns depends on the activities, and non--trivial time
correlations develop as the neuron temperature is lowered, namely, in region
O(I), where one even observes that $\tau _{\nu ,\zeta }\neq \tau _{\zeta
,\nu }.$ %
The behavior in O(I) suggests using our algorithm to code spatial--temporal
information \cite{lau}.

Summing up, we have illustrated several aspects of the behavior of
Hopfield--like neural automata. This consists of neurons and synapses that
evolve on the same time scale but subject to different thermal noises.
Furthermore, different network architectures have been considered. It
follows that a power--law topology, which is known to characterize many
natural, including neural systems is advantageous compared to the
corresponding diluted network. We also find definite evidence that \textit{%
hubs}, i.e. the few most highly connected nodes in a scale--free architecture,
play a fundamental role in making the retrieval of information more robust
and efficient. These findings, whose validity in natural systems is to be
checked, suggest paths for a convenient design of artificial systems. We
have also demonstrated that, for appropriate parameter values, neural
automata perform much more efficiently when one lets the subsystem of
synapses to constantly and coherently visit all the stored patterns in a
convenient way. This is fully consistent with two main empirical 
observations, namely,
that memory is a global dynamic phenomenon, and that oscillations are
essential to cortex functions. The neural automata has also been shown to
exhibit in some situations spatial--temporal attractors, which may be
relevant to simple olfactory processing, for instance.
\begin{figure}[h!]
\begin{center}
\epsfxsize=6.5cm
\epsfbox{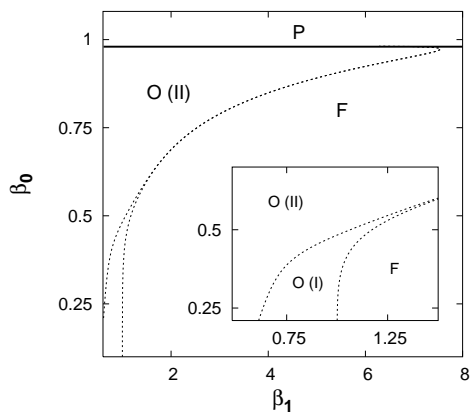}
\caption{Phase diagram for the FCN automata depicting
three coexistence curves that define several phases; see the main text.}
\end{center}
\end{figure}
We acknowledge financial support from MCyT--FEDER (project BFM2001-2841 and
a \textit{Ram\'{o}n y Cajal} contract), the EU COSIN project IST2001-33555,
and the UGR--MADOC agreement.


\begin{thebibliography}{99}
\bibitem{note1} It is worth remarking that this and several other details
here are rather arbitrary. They mainly intend an easy illustration of the
relevant result, and can generally be waived or modified ---some times
rather straightforwardly.
\end{thebibliography}
\end{document}